\documentclass[12pt]{article}
\usepackage{epsf}
\usepackage{amsmath}
\usepackage{amsfonts}
\usepackage{amssymb}
\usepackage{graphicx}
\usepackage{color}
\usepackage{psfrag}
\usepackage{cite}
\usepackage{rotating}
\usepackage{wrapfig,lipsum,booktabs}
\usepackage{lscape}
\usepackage{longtable}
\usepackage{amscd}

\usepackage{ifpdf}

\newcommand{\bmat}{\left(\begin{array}}
\newcommand{\emat}{\end{array}\right)}

\def\yzero{\smash{\hbox{$y\kern-4pt\raise1pt\hbox{${}^\circ$}$}}}

\def\be{\begin{equation}}
\def\ee{\end{equation}}
\def\bea{\begin{eqnarray}}
\def\eea{\end{eqnarray}}
\def\beq{\begin{equation}}
\def\eeq{\end{equation}}
\def\beqa{\begin{eqnarray}}
\def\eeqa{\end{eqnarray}}

\def\-{\hphantom{-}}

\def\s2{\frac{1}{\sqrt2}}

\def\beq{\begin{equation}}
\def\eeq{\end{equation}}
\def\beqa{\begin{eqnarray}}
\def\eeqa{\end{eqnarray}}

\def\IF{\relax{\rm I\kern-.18em F}}
\def\II{\relax{\rm I\kern-.18em I}}

\def\Dsl{\,\raise.15ex\hbox{/}\mkern-13.5mu D} 

\catcode`\@=11   
\newdimen\@rotdimen
\newbox\@rotbox  

\def\@vspec#1{\special{ps:#1}}
\def\@rotstart#1{\@vspec{gsave currentpoint currentpoint translate
   #1 neg exch neg exch translate}}
\def\@rotfinish{\@vspec{currentpoint grestore moveto}}
\def\@rotr#1{\@rotdimen=\ht#1\advance\@rotdimen by\dp#1%
   \hbox to\@rotdimen{\hskip\ht#1\vbox to\wd#1{\@rotstart{90 rotate}%
   \box#1\vss}\hss}\@rotfinish}
\def\@rotl#1{\@rotdimen=\ht#1\advance\@rotdimen by\dp#1%
   \hbox to\@rotdimen{\vbox to\wd#1{\vskip\wd#1\@rotstart{270 rotate}%
   \box#1\vss}\hss}\@rotfinish}%
\def\@rotu#1{\@rotdimen=\ht#1\advance\@rotdimen by\dp#1%
   \hbox to\wd#1{\hskip\wd#1\vbox to\@rotdimen{\vskip\@rotdimen
   \@rotstart{-1 dup scale}\box#1\vss}\hss}\@rotfinish}%
\def\@rotf#1{\hbox to\wd#1{\hskip\wd#1\@rotstart{-1 1 scale}%
   \box#1\hss}\@rotfinish}%
\def\rotate{\@ifnextchar[{\@rotate}{\@rotate[l]}}
\def\@rotate[#1]#2{\setbox\@rotbox=\hbox{#2}\@nameuse{@rot#1}\@rotbox}

\catcode`\@=12

\topmargin
-1.5cm
\textwidth
15.5cm
\textheight
23.5cm
\oddsidemargin
0.7cm
\evensidemargin
1.2cm

\begin{document}

\makeatletter
\@addtoreset{equation}{section}
\makeatother
\renewcommand{\theequation}{\thesection.\arabic{equation}}
\pagestyle{empty}
\begin{center}
\LARGE{Inference of ancestral recombination graphs through topological data analysis\\[10mm]}
\large{P.~G.~C\'amara${}^{1,2}$, A.~J.~Levine${}^2$ and R.~Rabad\'an${}^1$\\[6mm]}
\small{
${}^1$Department of Systems Biology and Department of Biomedical Informatics,\\
Columbia University College of Physicians and Surgeons,\\
1130 St. Nicholas Ave., New York \\[0.3em]
${}^2$The Simons Center for Systems Biology,\\
Institute for Advanced Study, Princeton, NJ 08540.
\\[1cm]}
\small{\bf Abstract} \\[0.5cm]
\end{center}

{\small
The recent explosion of genomic data has underscored the need for interpretable and comprehensive analyses that can capture complex phylogenetic relationships within and across species. Recombination, reassortment and horizontal gene transfer constitute examples of pervasive biological phenomena that cannot be captured by tree-like representations. Starting from hundreds of genomes, we are interested in the reconstruction of potential evolutionary histories leading to the observed data. Ancestral recombination graphs represent potential histories that explicitly accommodate recombination and mutation events across orthologous genomes. However, they are computationally costly to reconstruct, usually being infeasible for more than few tens of genomes. Recently, Topological Data Analysis (TDA) methods have been proposed as robust and scalable methods that can capture the genetic scale and frequency of recombination. We build upon previous TDA developments for detecting and quantifying recombination, and present a novel framework that can be applied to hundreds of genomes and can be interpreted in terms of minimal histories of mutation and recombination events, quantifying the scales and identifying the genomic locations of recombinations. We implement this framework in a software package, called TARGet, and apply it to several examples, including small migration between different populations, human recombination, and horizontal evolution in finches inhabiting the Gal\'apagos Islands.  
 }

\newpage
\setcounter{page}{1}
\pagestyle{plain}
\renewcommand{\thefootnote}{\arabic{footnote}}
\setcounter{footnote}{0}

\tableofcontents

\section{Introduction}
\label{sec1}
Since the publication of the first draft of the human genome \cite{lander2001initial, venter2001sequence}, there has been an explosion in genomic data. The genomes of thousands of different human individuals have been sequenced \cite{10002012integrated}, several hundreds of eukaryotic genomes have been characterized, and new viral, bacterial and archaeal species are being sequenced on an almost daily basis \cite{edwards2005viral, albertsen2013genome}. Darwin provided a historical dimension to the taxonomical enterprise, proposing that closely related species in the hierarchical taxonomy share ancestors. Since then, tree-like structures have been proposed to represent the evolutionary/historical relationship between organisms. In the last few years, however, the richer and more comprehensive genomic characterization of many organisms have underscored the need of representations that are not strictly tree-like. Phenomena such as horizontal gene transfer in bacteria \cite{ochman2000lateral}, the ability of viruses to borrow and lend genes across species, and hybridization in metazoa (in plants, in particular \cite{burke2001genetics, adams2005polyploidy}) are exposing some of the limitations imposed by tree-like phylogenetic structures. The definition of species itself becomes cumbersome in bacteria and viruses \cite{doolittle2006genomics}. Within many species, including humans, genetic recombination is so pervasive that tree-like representations are useless. It is then natural to wonder what other frameworks could be used to capture phylogenetic relationships without losing the interpretability and simplicity of trees \cite{doolittle1999phylogenetic, woese2004new, o2011stands}. Of particular interest are representations that reduce to trees when evolution is tree-like; that capture genetic relations between ancestors, and identify genomic regions originating from different ancestral lineages; and, more generally, that allow for an interpretation of the observed data in terms of a chronological sequence of events.

Several such frameworks have been proposed in the last two decades. The study of phylogenetic networks has been an area particularly active \cite{huson2010phylogenetic, huson2011survey, morrison2011introduction}. Phylogenetic networks provide representations that extend trees to graphs (networks), generating loops when the data does not fit into a tree. Some of those methods can easily be applied to more than one hundred genomes \cite{bandelt1992split, bandelt1995mitochondrial, huson1998splitstree, bandelt1999median, bandelt2000median, huson2006application} providing the opportunity for large-scale representations. However, the biological interpretation of these representations is limited, as loops represent inconsistencies with trees, but it is unclear how these inconsistencies arose historically, what genomic regions were involved, or how frequently an exchange happened. Other types of representations, sometimes named \emph{explicit networks}~\cite{huson2010phylogenetic, gusfield2014recombinatorics}, do aim to provide a historical account in terms of a chronology of events. Ancestral recombination graphs (ARGs) provide potential explanations of the observed data in terms of a progression of recombination and mutation events. As in trees, mutations are represented as events along the branches. Recombinations, however, appear as the fusion of two parental branches into one offspring branch. ARGs provide simple histories that can be used in association mapping \cite{minichiello2006mapping, sutter2007single, wu2008association}, SNP genotyping \cite{le2011snp} or inference of the frequency and scale of recombination \cite{wall2000comparison}. However, these applications are hindered by the computational infeasibility of constructing ARGs that explain hundreds of sequences. The construction of minimal ARGs, containing the minimum number of recombination events required to explain the sample in absence of convergent evolution and back-mutation, is an NP-hard problem \cite{wang2001perfect, bordewich2005computational, bordewich2007computing}. Several approximations have been developed in the last few years, including galled trees \cite{gusfield2004optimal, gusfield2005optimal}, branch and bound \cite{song2005efficient}, heuristic \cite{minichiello2006mapping} and sequentially Markov coalescent approaches \cite{rasmussen2014genome}.

\begin{figure}[ht!]
\centering
\includegraphics[scale=0.7]{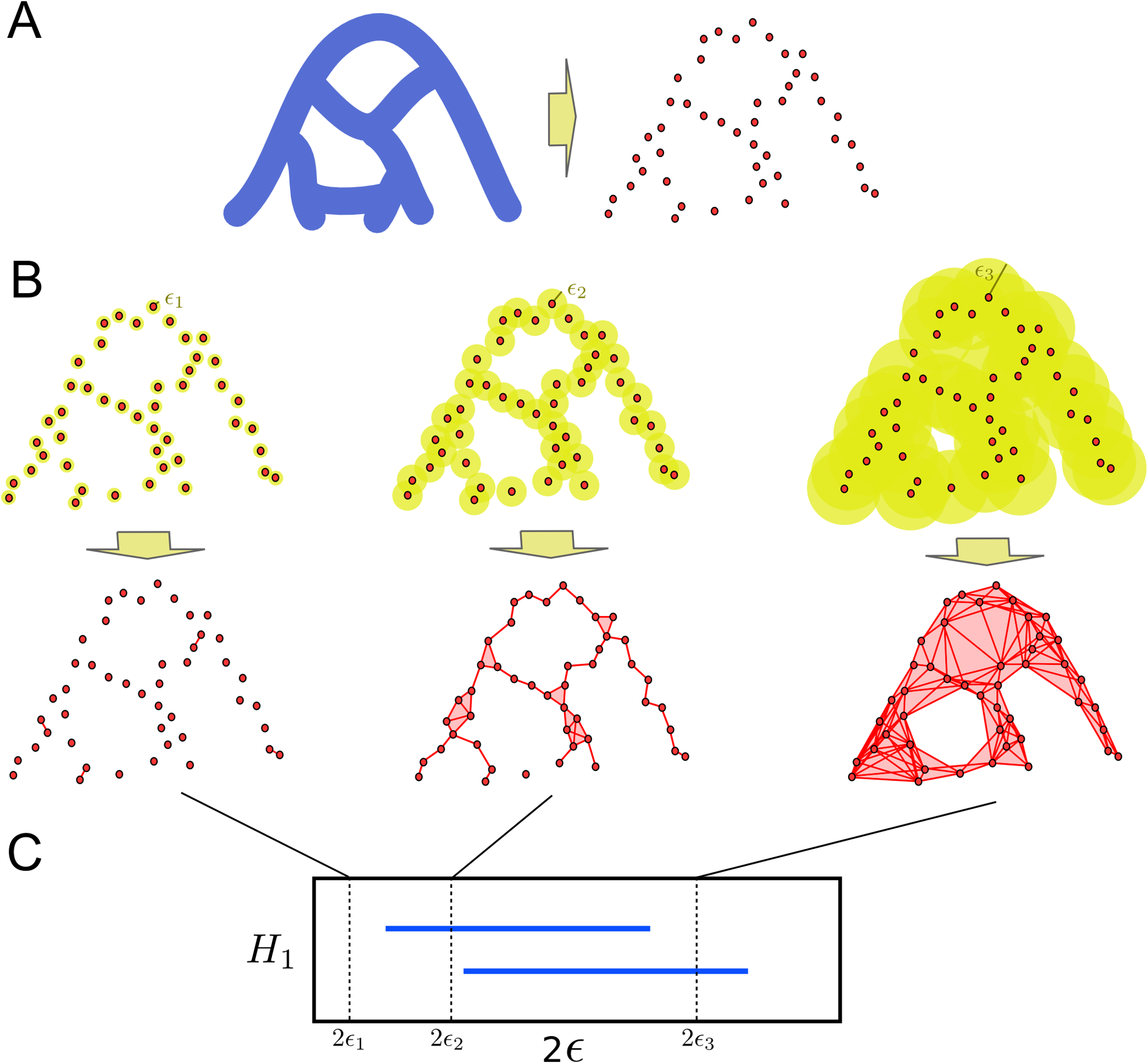}
\caption{{\bf Topological data analysis}. (A) Topological data analysis aims to infer the topological features (e.g. loops, voids, etc.) of an unknown space from a finite set of sampled points. (B) Persistent homology, a tool of TDA, builds simplicial complexes (generalizations of networks that include higher dimensional elements like triangles and tetraheadra), by taking balls of radius $\epsilon$ centred on the sampled points. Points are connected in the simplicial complex if the corresponding balls intersect. This construction is known as Vitoris-Rips complex. Persistent homology tracks how the topological features of Vietoris-Rips complexes change with $\epsilon$. (C) Barcodes are suitable representations of persistent homology. Each interval in the barcode represents the range of $\epsilon$ across which a particular topological feature (for instance, a loop) is present in the inferred topology. In this figure, the barcode of the first persistent homology, that tracks the presence of loops, is shown. The two intervals in the barcode correspond to the two loops present in the original space.}
\label{fig0}
\end{figure}

Recently, a new framework to study genomic relationships has been proposed \cite{chan2013topology, emmett2014characterizing, emmett2014parametric}, based on topological data analysis \cite{carlsson2009topology, edelsbrunner2002topological, zomorodian2005computing}. Topology is the area of mathematics that aims to characterize properties of spaces up to continuous deformations, for instance the number of disconnected components, loops and holes of a space. TDA extends the concepts and tools of topology to finite metric spaces, that is, finite sets of points and distances between them. Taking the premise that a set of points has been sampled from an unknown underlying space, TDA attempts to infer the topological features of the space (Fig.~\ref{fig0}A). Stability results \cite{cohen2007stability, chazal2009gromov, chan2013topology} guarantee that small fluctuations in the data only create small changes in the inferred topological features, providing robust characterizations of the data. 

In a TDA framework, genomes are characterized by points in a high dimensional space where pairwise distances are genetic distances between sequences. Assuming that each genomic site mutates at most once across the evolutionary history of the sample, the genetic distance between two genomes can only increase with the acquisition of novel mutations. The only way of ``closing'' a loop (a close path) in this space is therefore by means of a recombination event \cite{chan2013topology}. Hence, an approach to studying recombination in the sample of genetic sequences is to study the loops that those sequences generate when represented in the above way.

A valuable attribute of TDA methods is that they are informative about the scale or size of the inferred topological features. Given a finite set of data points, there is an infinite number of spaces that are compatible with the points. TDA structures this spectrum of possibilities by introducing a notion of scale (Fig.~\ref{fig0}B): at a given scale $\epsilon$, two points are connected in the underlying space if their distance is smaller than $\epsilon$. Topological features compatible with the data can be then summarized in terms of sets of intervals, named \emph{barcodes}~\cite{ghrist2008barcodes}  (Fig.~\ref{fig0}C). Each interval in a barcode represents the range of scales across which a particular topological feature (e.g. a loop) is present in the inferred topological space. In the genomic context introduced above, barcodes of loops summarize the frequency and scale (mutational distance between recombining sequences) of recombination events, and provide a basic structure on which statistics of genomic exchange can be built \cite{emmett2014parametric}.

TDA methods are particularly well suited for large datasets. In the context of molecular phylogenetics and evolution, they have been applied to the study of viral recombination and reassortment \cite{chan2013topology}, bacterial species \cite{emmett2014characterizing} and point estimators in population genetics \cite{emmett2014parametric}. However, these implementations of TDA have limitations, as they are not tailored for the biological problem they try to address. Specifically, traditional TDA methods only use information about genetic distances between sequences, and so they discard the full structure of segregating characters, missing numerous recombination events that are required to explain the data. Relatedly, it is unclear which specific evolutionary histories explaining the data TDA informs about, and what is the precise relation between barcodes and these histories.

Here we address these two important aspects, improving on the scalable capabilities of TDA to extract robust information on the possible evolutionary histories of a sample of genetic sequences. In particular, we show that by systematically sampling subsets of segregating sites and performing TDA,  we are able to identify most of the necessary recombination events identified by bound methods \cite{hudson1985statistical, myers2003bounds, song2005efficient}, providing a significant improvement of past methods \cite{chan2013topology, emmett2014characterizing, emmett2014parametric} in terms of interpretation and sensitivity. Moreover, we introduce a novel type of graph (\emph{topological ARG} or \emph{tARG}), closely related to minimal ARGs, that captures ensembles of minimal recombination histories; and we show that TDA informs about the topological features and genetic scales of these graphs. Like minimal ARGs \cite{minichiello2006mapping, gusfield2014recombinatorics}, tARGs can be considered as explicit, parsimonious, interpretable phylogenetic representations. The main advantage of tARGs and barcodes versus minimal ARGs is, however, the possibility of obtaining such phylogenetic information in polynomial time, which allows us to deal with hundreds of sequences. We have implemented this method in a software, called \texttt{TARGet}, and have illustrated it with several examples, including small migration between diverging populations, human recombination, and horizontal evolution of finches inhabiting the Gal\'apagos archipelago. The software, instructions and example files used in the manuscript can be obtained from  \texttt{https://github.com/RabadanLab/TARGet}.

\section{Results}
\subsection{Topological ARGs}
An ARG is an explicit phylogenetic network representing a possible evolutionary history of a sample of genetic sequences, where only mutation and recombination events are present and convergent evolution is not considered and so never occurs \cite{griffiths1996ancestral, griffiths1997ancestral, gusfield2014recombinatorics}. ARGs are very useful constructs in population genetics and phylogenetics. However, the problem of building a minimal ARG from a set of genetic sequences is known to be NP-hard \cite{wang2001perfect, bordewich2005computational, bordewich2007computing}. The use of ARGs has therefore been traditionally limited to small samples, consisting of a handful of sequences. 

In this section, we introduce a particular class of minimal ARGs and a set of related graphs. Then, using computational algebraic topology, in the next section we show that it is possible to extract, in polynomial time, phylogenetic information from this class of minimal ARGs, without having to explicitly construct them. Thus, by restricting to this specific class of graphs, we are able to extend the realm of ARGs to large samples of sequences.

To be specific, we consider a sample $\mathcal{S}$ consisting of $n$ distinct genetic sequences with $m$ binary segregating characters. The latter can be single nucleotide polymorphisms (SNPs), indels, gene duplications or any other genetic trait that takes one of two possible states, 0 or 1, in each sequence. An ARG is then formally defined as a directed acyclic graph $\mathcal{N}$ with $n$ leaf nodes and a unique root node, where every node other than the root has in-degree one (tree node) or two (recombination node), every segregating character labels a unique edge in $\mathcal{N}$ (infinite sites assumption), and every sequence in  $\mathcal{S}$ labels a unique leaf in $\mathcal{N}$. Moreover, each node in $\mathcal{N}$ is labelled by a $m$-length binary sequence, such that the sequence labelling a tree node differs from the sequence of the parent node only at the character labelling the edge that connects the two nodes; and the sequence labelling a recombination node is a combination of the sequences labelling the two parent nodes. Single-crossover recombinant sequences are formed by taking the first $k$ sites from the sequence of one of the parent nodes (\emph{prefix}) and appending the last $m-k$ sites from the sequence of the other parent node (\emph{suffix}), for $k\in[1,m-1]$.

There is an infinite number of ARGs that can explain a given sample $\mathcal{S}$ \cite{gusfield2014recombinatorics}. A stochastic model, such as the coalescent model with recombination \cite{hudson1983properties, griffiths1996ancestral}, would assign probabilities to each possible ARG. Here, however, we adopt a parsimony approach and consider ARGs that are minimal (in a sense defined below), without assuming an underlying probabilistic model. Such a model-independent approach has proven useful in summarizing genetic sequences into evolutionary histories where all events are required. 

\begin{figure}[ht!]
\centering
\includegraphics[scale=0.8]{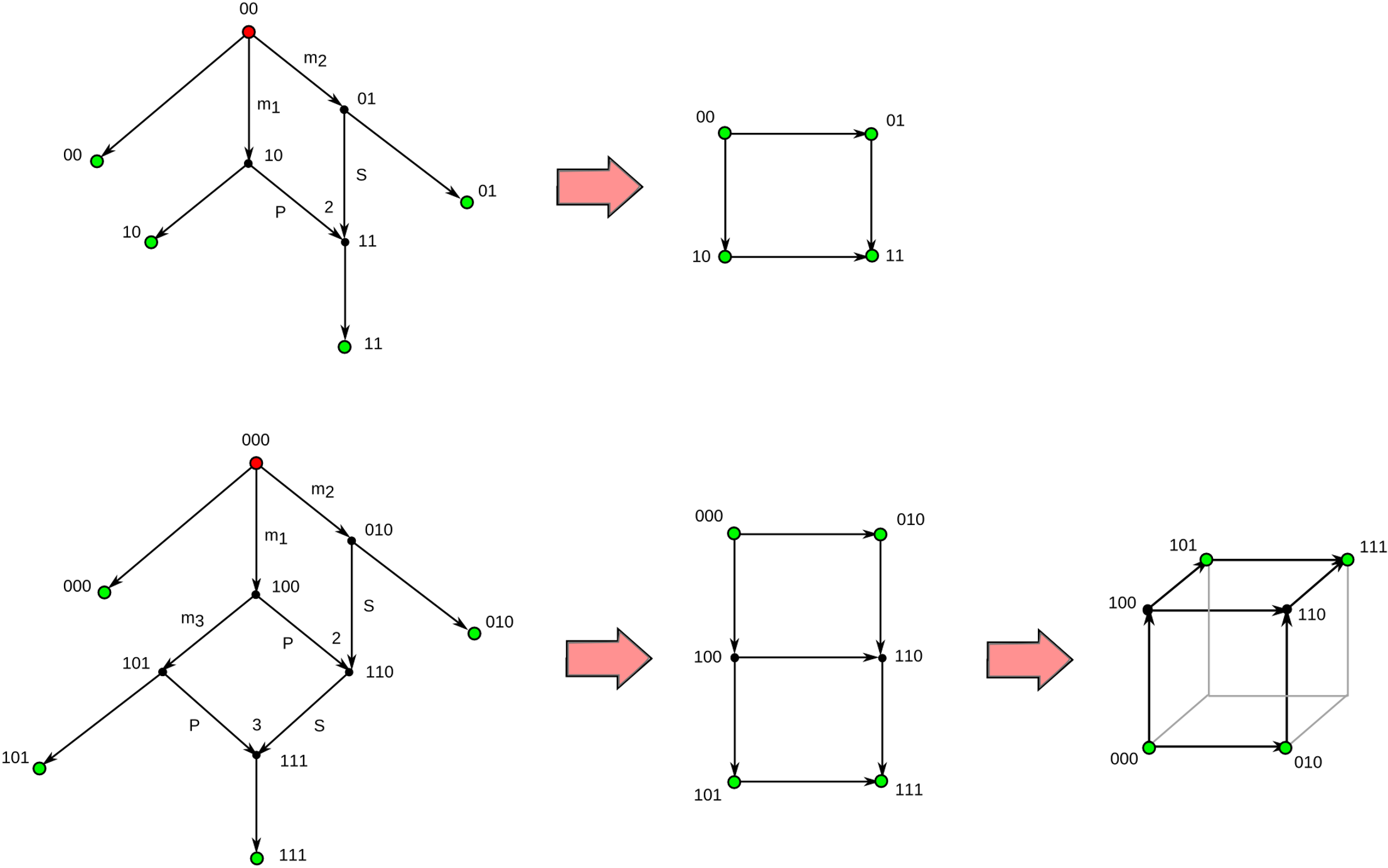}
\caption{{\bf ARGs and condensed graphs}. Two examples of ultra-minimal ARGs and the condensed graphs that result from collapsing their unlabelled edges. The root node is marked red whereas sampled nodes are marked green. Mutations in the $r$-th character are indicated by $m_r$. Edges pointing to a recombination node are labelled with the letter P or S, depending on whether they contribute to the prefix or suffix of the recombinant sequence. Recombinant nodes are marked with the position of the recombination breakpoint. All nodes are labelled by their sequence of characters. Condensed graphs of ARGs can be embedded into $m$-dimensional hypercubes and their diagonals.}
\label{fig1}
\end{figure}

\begin{figure}[ht!]
\centering
\includegraphics[scale=0.8]{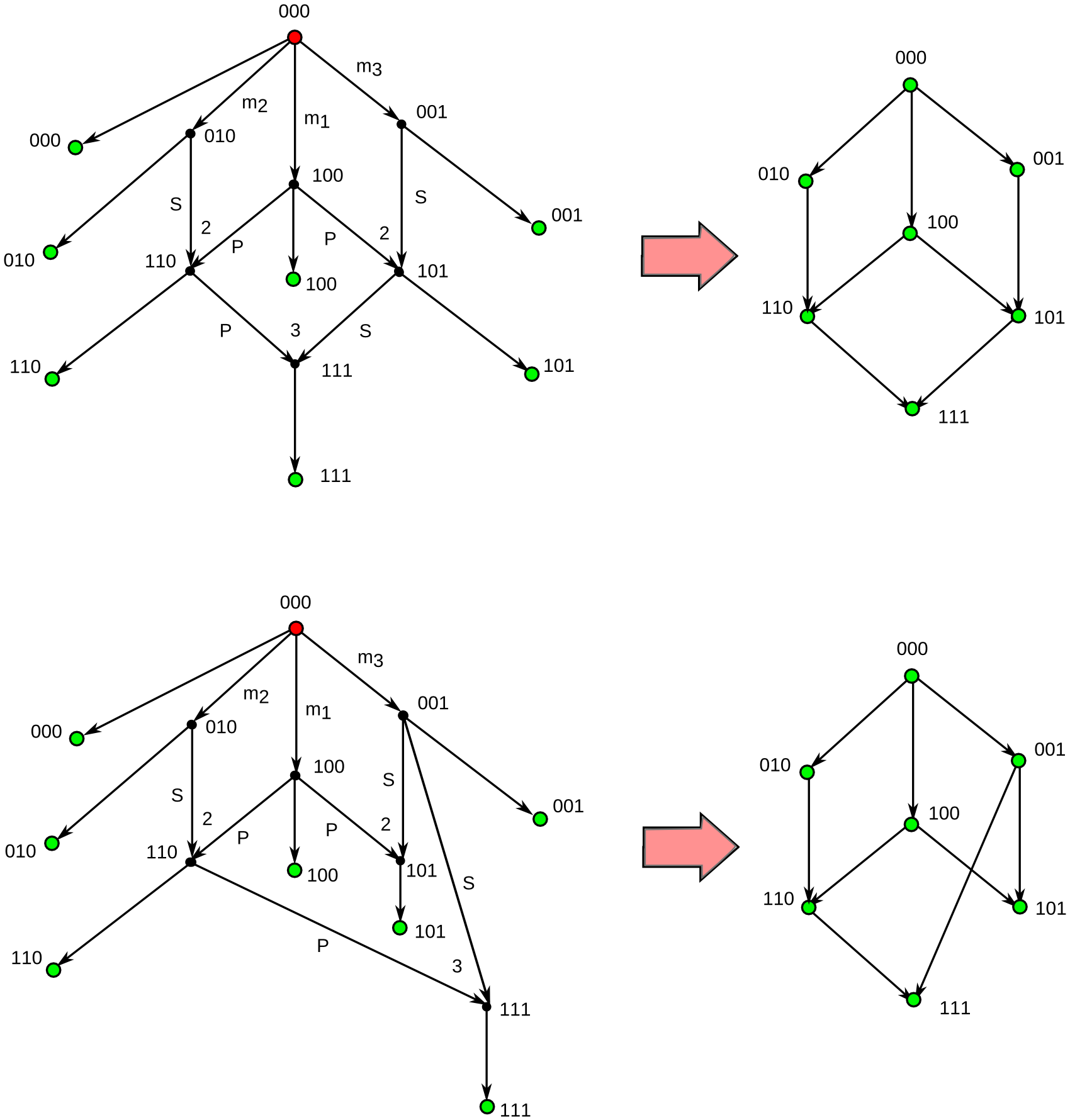}
\caption{{\bf Ultra-minimal ARGs}. Two examples of ARGs containing the minimum number of recombination events, $R_{\rm min}=3$, required to explain a sample of $n=7$ sequences with $m=3$ segregating sites. Both ARGs are minimal ARGs. However, only the minimal ARG at the bottom is an ultra-minimal ARG.}
\label{fig2}
\end{figure}

Specifically, we consider ARGs that contain exactly the minimum number $R_{\rm min}$ of single-crossover recombinations required to explain the sample, and that minimize the function
\begin{equation}
\label{eq1}
D(\mathcal{N}) = \sum_{r=0}^{R_{\rm min}}d_{r}
\end{equation}
where the sum runs over all recombination events in $\mathcal{N}$, and $d_r$ is the Hamming distance between the two parental sequences involved in the $r$-th recombination. This is a more restricted definition of minimal ARG than the one that usually appears in population genetics literature \cite{gusfield2014recombinatorics}, where the condition on $D(\mathcal{N})$ is generally not required. We use the term \emph{ultra-minimal ARG} to refer to this restricted type of minimal ARG. Ultra-minimal ARGs are thus minimal ARGs where recombination events involve parental sequences that are as genetically close as possible. They introduce a higher level of parsimony than minimal ARGs, being informative not only about the minimum number of recombination events, but also about the minimum genetic distance between the recombining sequences that took part in those events. By construction, an ultra-minimal ARG explaining any given sample always exists. Examples are shown in Figs.~\ref{fig1} and \ref{fig2}.

A minimal ARG can be condensed by collapsing all unlabelled edges, so that the resulting graph can be embedded into an $m$-dimensional hypercube and its diagonals (that is, the line segments joining non-consecutive vertices) (Fig.~\ref{fig1}). The number of edges and vertices of such a condensed representation is $m+2R_{\rm min}$ and $m+R_{\rm min}+1$, respectively, whereas the number of independent loops is $R_{\rm min}$, where a loop is said to be independent if it cannot be embedded in the union of other loops. In this representation, the distance between two nodes is defined as the number of edges in the shortest path connecting the nodes, and is equal to the Hamming distance between the corresponding sequences.

\begin{figure}[ht!]
\centering
\includegraphics[scale=0.8]{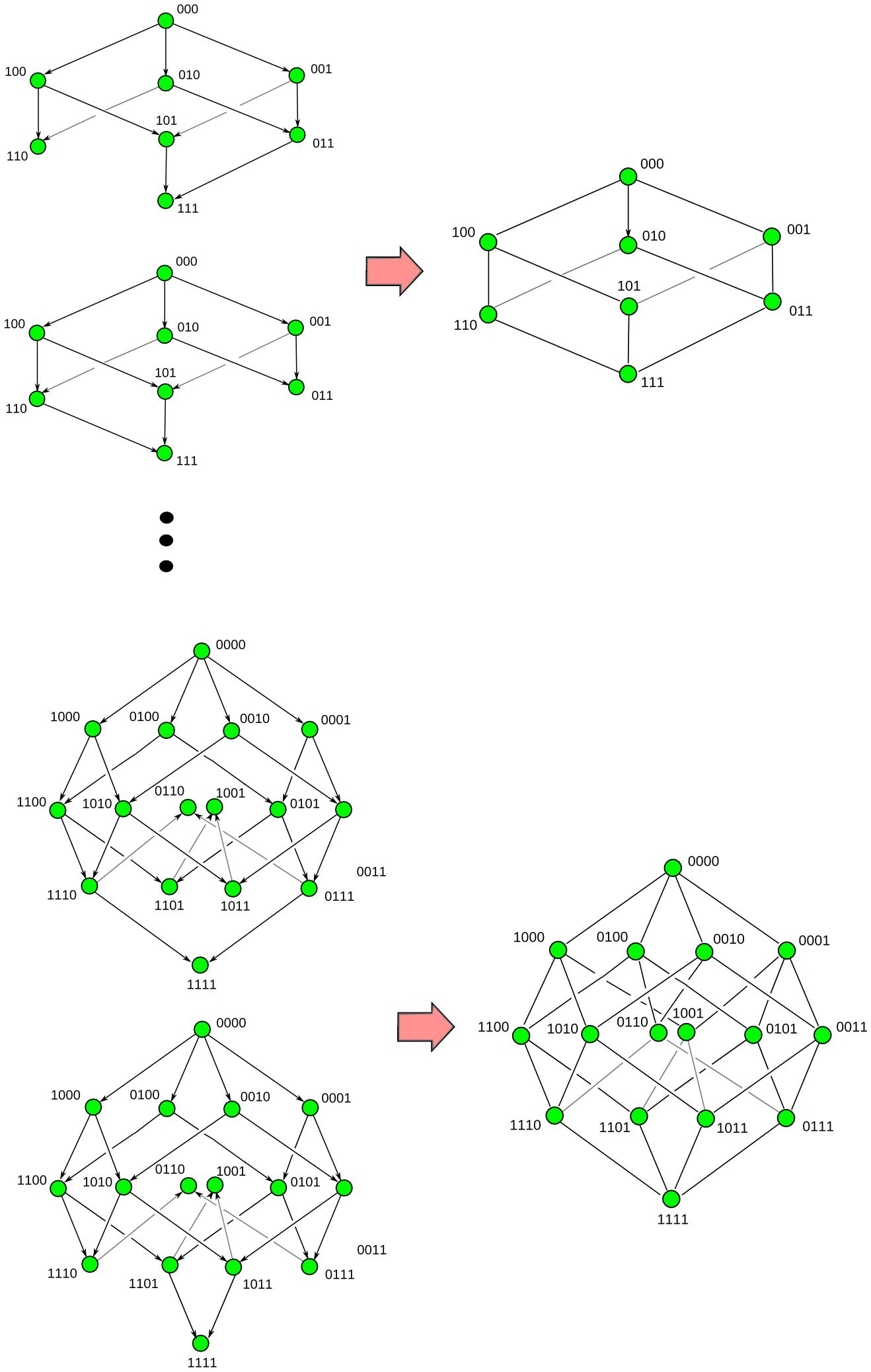}
\caption{{\bf Topological ARGs}. Examples of condensed ultra-minimal ARGs (left) and their corresponding tARGs (right). In a tARG the edges are completely determined by the vertices. The topology of the resulting tARG can differ from that of the original condensed ultra-minimal ARGs.} 
\label{fig3}
\end{figure}

Given a sample $\mathcal{S}$ of genetic sequences, we would like to obtain information about the ultra-minimal ARGs that explain $\mathcal{S}$, without explicitly constructing them. To that end, we consider the undirected graph $\mathcal{G}=(V,E)$, with vertices $V$ and edges $E=E_1\cup\ldots\cup E_l$, that results from the union of all condensed ultra-minimal ARGs $\mathcal{G}_i=(V, E_i)$ explaining $\mathcal{S}$ and having the same set of vertices $V$ (Fig.~\ref{fig3}). We call this construction \emph{topological ARG} (tARG). A tARG therefore summarizes the collection of most parsimonious histories associated to a sample of genetic sequences. However, unlike minimal ARGs, tARGs are completely determined by their vertices. By considering tARGs instead of minimal ARGs, we are able to reduce an NP-hard problem into a much simpler (but still very informative) topological problem, as we describe in next section. 

\subsection{Persistent homology and recombination inference}
Topological data analysis has emerged during the last decade as a branch of applied topology that attempts to infer topological features of spaces (such as the number of loops and holes) from sets of sampled points \cite{carlsson2009topology}. The topological features of a space are preserved under continuous deformations of the space and can be arranged in mathematical structures called \emph{homology groups} \cite{hatcher2002algebraic}. We refer the reader to  refs.~\cite{hatcher2002algebraic, ghrist2014elementary} for formal definitions and basic introductions to algebraic topology. In brief, the $n^\textrm{th}$ homology group of a space is an algebraic structure that encompasses all $(n+1)$-dimensional holes of the space. Of special interest to us is the first homology group, whose elements correspond to loops. 

Homology groups can be computed by replacing the original space with a simpler one, known as \emph{simplicial complex}, which has the same topological
features as the original space but consists of a finite set of elements (Fig.~\ref{fig0}B). A simplicial complex is a generalization of a network that, in addition to nodes
and vertices, includes higher dimensional elements like triangles and tetrahedra.
Simplicial complexes are powerful because they allow the implementation of algebraic
operations to extract the topological features of the space. 

When only a finite set of points of the space is given, there is still a well-defined notion of homology groups, known as \emph{persistent homology} \cite{edelsbrunner2002topological, zomorodian2005computing}, which capture the topological features of the underlying space. At each value of a scale parameter $\epsilon$, a simplicial complex (known as Vietoris-Rips complex) can be constructed by considering the intersections of balls of radius $\epsilon$ centred at the sampled points (Fig.~\ref{fig0}B). Points are joined if their corresponding balls intersect. This process produces a sequence of simplicial complexes parametrized by $\epsilon$, from which persistent homology can be computed using available algorithms \cite{edelsbrunner2002topological, zomorodian2005computing}. Remarkably, the computation time of persistent homology is polynomial in the number of points \cite{edelsbrunner2002topological, zomorodian2005computing}.

Persistent homology can be represented using barcodes~\cite{ghrist2008barcodes}. These are graphical representations where each element of persistent homology is represented by a segment spanning the interval $[\epsilon_b,\, \epsilon_d]$, where $\epsilon_b$ and $\epsilon_d$ are the values of the parameter $\epsilon$ at which the corresponding feature is respectively formed and destroyed in the sequence of simplicial complexes (Fig.~\ref{fig0}C). Thus, each segment in a barcode represents a topological feature inferred from the data, and the position and length of the segment are informative of the size of the topological feature. The values $\epsilon_b$ and $\epsilon_d$ are referred as \emph{birth} and \emph{death time} of the topological feature, respectively.

\begin{figure}[ht!]
\centering
\includegraphics[scale=0.8]{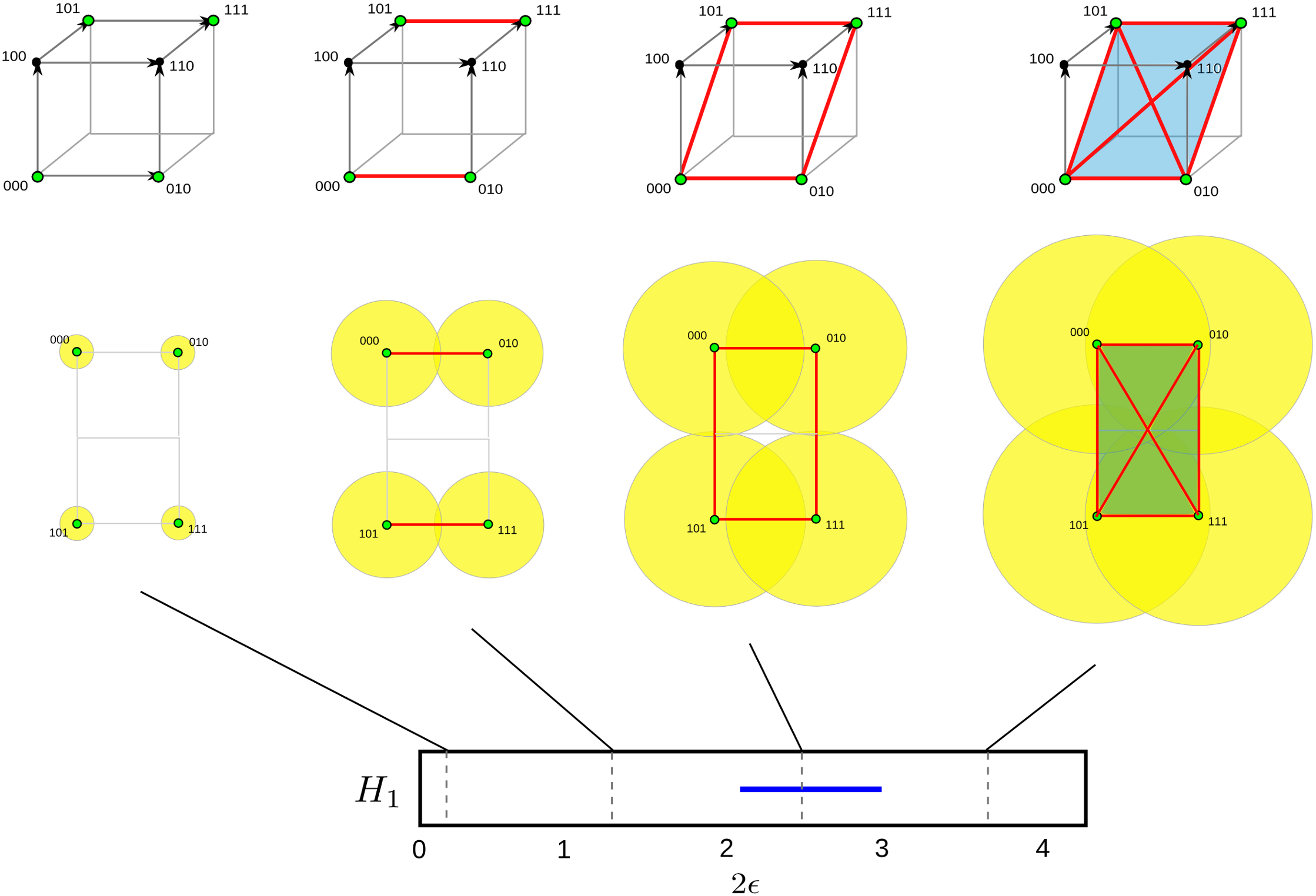}
\caption{{\bf Persistent homology of a sample of genetic sequences}. Barcode and Vietoris-Rips complexes at several values of the parameter $\epsilon$, for the sample of sequences $\mathcal{S} = \{000,\, 010,\, 101,\, 111\}$. Only the first homology group ($H_1$) is shown. At small $\epsilon$ the four sampled points are disconnected. Increasing $\epsilon$ leads to a loop, that appears as a single element of $H_1$. Further increasing $\epsilon$ fills in the loop, leading to a single connected surface. An ultra-minimal ARG explaining $\mathcal{S}$, and the corresponding tARG are shown in Fig.~\ref{fig1} (bottom). The barcode only captures one of the $R_{\rm min}=2$ recombination events.} 
\label{fig4}
\end{figure}

In the current context, we exploit the use of persistent homology to infer topological features of an unknown tARG, given a set of sampled nodes (Fig.~\ref{fig4}). The use of persistent homology to detect the presence of recombination in genetic samples was proposed in \cite{chan2013topology}. However, the relation between persistent homology and explicit evolutionary histories incorporating recombination events was not studied. Our aim is inferring information about the loops of the tARG, as they correspond to recombination events present in the collection of most parsimonious histories explaining the sample. To that end, we consider the Hamming distance matrix of the sample and compute persistent homology using the algorithm developed in ref.~\cite{edelsbrunner2002topological, zomorodian2005computing}. Since computing the distance matrix and persistent homology requires respectively $\mathcal{O}(n^2m)$ and $\mathcal{O}(n^3)$ operations \cite{edelsbrunner2002topological, zomorodian2005computing}, the running time grows at most cubically with the number of genetic sequences. An advantage of using persistent homology instead of just counting loops in a nearest neighbour graph is that we also obtain valuable information about the genetic distances between recombining sequences. 

The barcode that results from this computation contains information about the number and size of the loops in the tARG underlying the sample (Fig.~\ref{fig4}). Each segment in the barcode represents a loop in the tARG, and therefore a recombination event in an ultra-minimal ARG explaining the sample. The position of each segment provides information about the genetic scales involved in the corresponding recombination event. Specifically, $2\epsilon_d$ sets an upper bound to the mutational distance between the two recombining sequences, since all pairwise distances between nodes in the loop are smaller than $2\epsilon_d$. The number of segments in the barcode (namely, the dimension of the first persistent homology group) or \emph{persistent first Betti number}, $b_1$, is hence a lower bound of the number of recombination events in the tARG, $\overline{R}_{\rm min}$. Note that, since a tARG is the union of multiple minimal histories, $\overline{R}_{\rm min}$ can be larger than $R_{\rm min}$. In particular, $\overline{R}_{\rm min} > R_{\rm min}$ when there are three characters for which all eight possible allele combinations appear in the sample. In general, this can only happen at very large recombination rates.

\subsection{The barcode ensemble of a sample}

The sensitivity of persistent homology to detect recombination decreases as the number $m$ of segregating characters increases. Indeed, in that case the dimensionality of the ambient space is larger and the sample becomes sparser. For this reason, $b_1$ is in general a loose lower bound of $\overline{R}_{\rm min}$. To address a similar problem, Myers and Griffiths introduced the idea of combining the local bounds that result from partitioning the sequence, building a more stringent global bound \cite{myers2003bounds}. In this way, information about the ordering of characters is incorporated and the location of recombination breakpoints is constrained in the sequence. This general idea was applied in \cite{myers2003bounds} to the haplotype bound, $n-m-1 \leq R_{\rm min}$, to built a stronger lower bound of $R_{\rm min}$, denoted $R_{\rm MG}$.

\begin{figure}[ht!]
\centering
\includegraphics[scale=0.8]{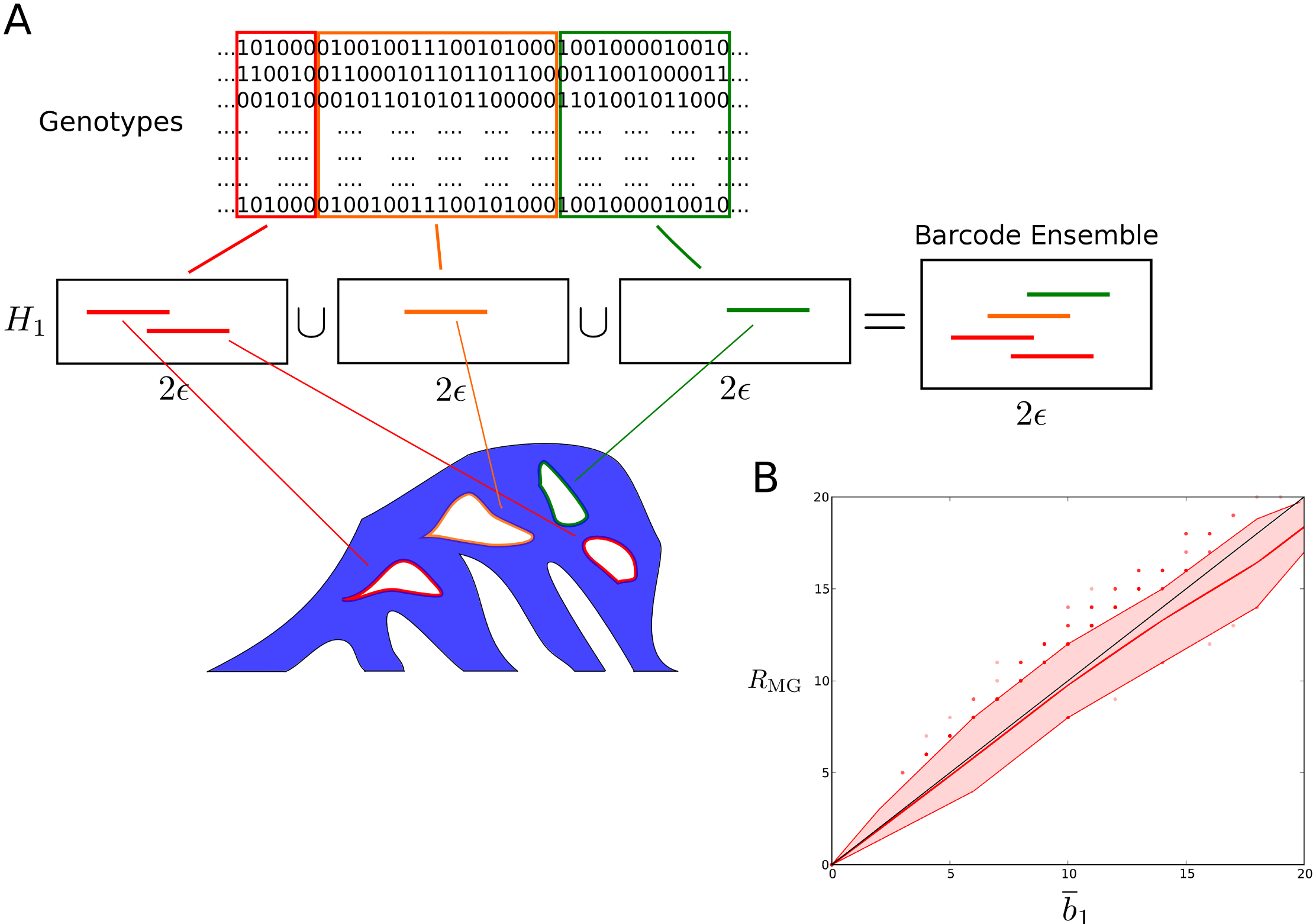}
\caption{{\bf Barcode ensemble of a sample}. (A) Schematic representation of the barcode ensemble of a genomic sample. Persistent homology is computed for each genomic interval of a partition of the sequence. Barcodes associated to different genomic intervals capture different recombination events. The union of all barcodes is the barcode ensemble. The total number of intervals in the barcode ensemble is denoted as ${\bar b_1}$. The partition is chosen such that ${\bar b_1}$ is maximized. (B) Comparison between lower bounds ${\bar b_1}\leq \overline{R}_{\rm min}$ and $R_{\rm MG}\leq R_{\rm min}$ in coalescent simulations. Values of $\bar b_1$ and $R_{\rm MG}$ for simulated samples of 40 sequences with 12 segregating sites, sampled from a population under the coalescent model with recombination. 4,000 samples were simulated in total. The colored band represents the interdecile range, whereas the central line represents the mean. The values of ${\bar b_1}$ and $R_{\rm MG}$ are strongly correlated (Pearson's $r=0.98$, $p<10^{-100}$). At very high recombination rates, $\bar b_1$ tends to be larger than $R_{\rm MG}$, as cases where $\overline{R}_{\rm min}>R_{\rm min}$ occur more frequently.}
\label{fig5}
\end{figure}

A similar idea can be applied in the context of barcodes to build a \emph{barcode ensemble}, given by the disjoint union of the persistent first-homology barcodes of a set of optimally chosen, non-overlapping intervals within the sequence alignment (Fig.~\ref{fig5}A). Given a partition of a genetic sequence, the barcode associated to each interval captures information about recombination events with breakpoint in that interval. Due to the curse of dimensionality mentioned in the previous paragraph, the union of the barcodes associated to two contiguous genomic intervals often captures more recombination events than the barcode associated to the union of the two genomic intervals. Therefore, by systematically exploring all possible partitions of the genetic sequence, it is possible to find a partition that maximizes the total number of bars in the barcodes. The solution is often not unique, as different partitions may lead to the same total number of bars. One may reduce this degeneration by considering additional criteria, such as also maximizing the total length of the bars (so that they are more informative about genetic distances). The formal details of the barcode ensemble construction are presented in the Methods section. 

The barcode ensemble incorporates information about the full structure of characters in the sample, largely increasing the sensitivity of persistent homology to recombination and providing information on the location of the recombination breakpoints in the sequence. The number of bars in the barcode ensemble, $\bar b_1$, is an improved lower bound of $\overline{R}_{\rm min}$, in the same way as $R_{\rm MG}$ is an improved lower bound of $R_{\rm min}$:

\begin{equation*}
\begin{CD}
\textrm{tARG} \ @>>> \ \overline R_{\rm min} \ \geq \ \bar b_1 \ \geq \ b_1 \\
@AAA\\
\textrm{(ultra) minimal ARG} \ @>>> \ R_{\rm min} \ \geq \ R_{\rm MG} \ \geq \ n-m-1 
\end{CD}
\end{equation*}

In biological data, $\bar b_1$ and $R_{\rm MG}$ are in general very close to each other (Fig.~\ref{fig5}B), as tARGs with $\overline{R}_{\rm min}>R_{\rm min}$ occur very rarely. However, unlike $R_{\rm MG}$, barcode ensembles provide additional phylogenetic information, such as bounds on the mutational distances between recombining sequences (note that birth and death times in barcode ensembles refer to local genetic distances, namely mutational distances across the genomic interval associated to the particular bar). These features put barcode ensembles at the very interesting interface between the fast, but phylogenetically limited, existing lower bounds to $R_{\rm min}$; and the slow, but phylogenetically rich methods for reconstructing minimal ARGs. We have implemented the computation of barcode ensembles in publicly available software, called \texttt{TARGet}.

\subsection{Examples}
We consider five examples that illustrate how the formal developments presented in previous sections can be used to extract useful phylogenetic information from samples of genetic sequences. The first example is a simple toy model where an explicit minimal ARG can be easily constructed. It displays how the information contained in the barcode ensemble of the sample directly maps to features of ultra-minimal ARGs. The second example, based on simulated data of two sexually reproducing populations exchanging genetic material at low rate, shows the applicability of persistent homology to large datatsets, consisting of several hundreds of sequences. It also demonstrates the use of phylogenetic information contained in the barcode ensemble to distinguish among various biological settings with similar recombination rates. The third and fourth examples consist respectively of 250 and 100 kilobase regions in the HLA and MS32 loci of $\sim 100$ humans, where several meiotic recombination hotspots localize. The fifth example consists of a 9 megabase scaffold in the genome of 112 Darwin's finches \cite{lamichhaney2015evolution}. These last three examples serve to illustrate the applicability of barcode ensembles to real datasets.

\paragraph{A simple example.}

\begin{figure}[ht!]
\centering
\includegraphics[scale=0.8]{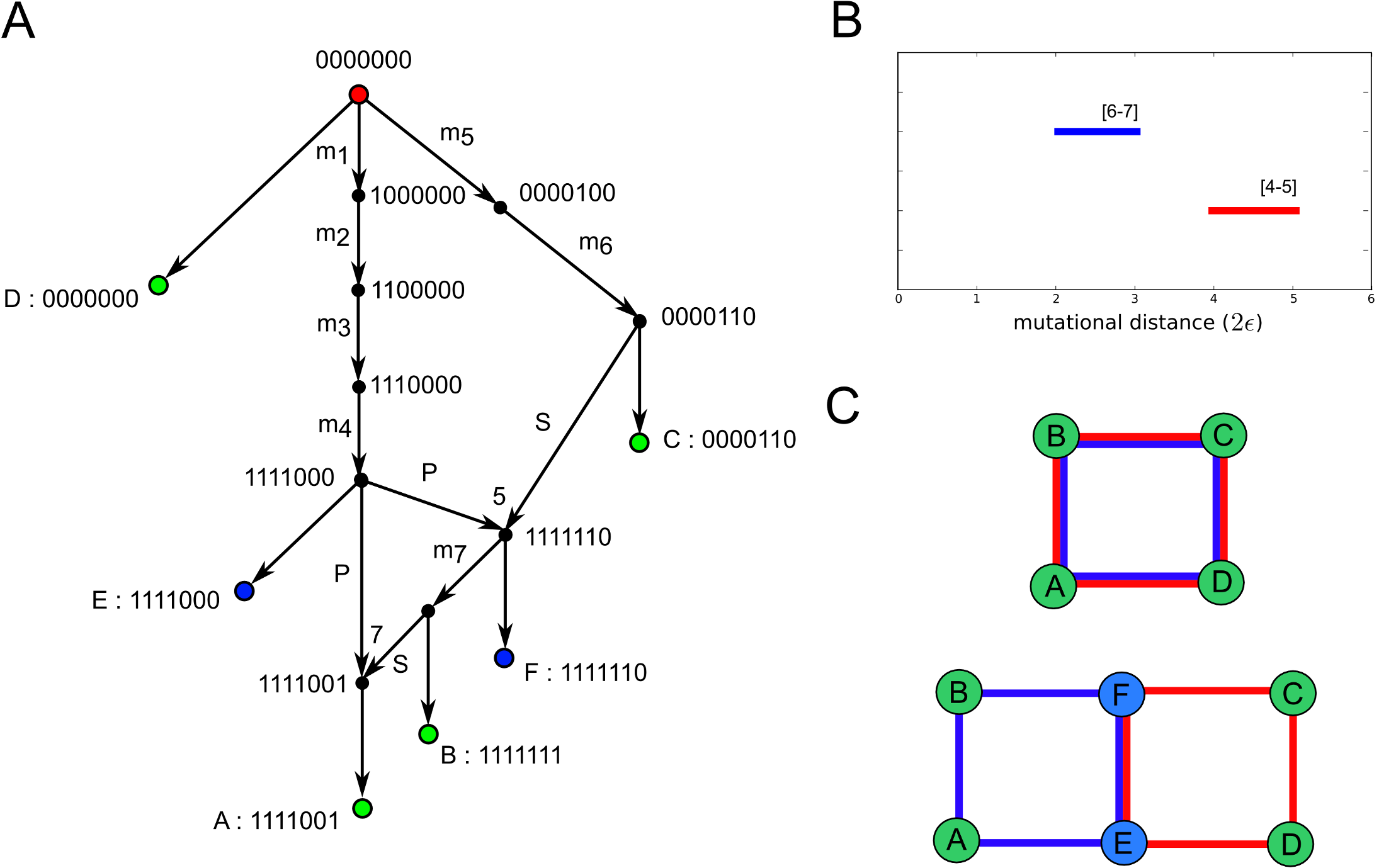}
\caption{{\bf Ultra-minimal ARG, first-homology barcode ensemble and reconstructed tARG of a sample of 4 sequences}. The four sampled sequences are represented by green leaf nodes in the ultra-minimal ARG depicted in (A). The ARG involves two single-crossover recombination events. Both recombination events and their genetic scales (mutational distance between recombining sequences) are correctly captured by the barcode ensemble of the samples, shown in (B). Intervals containing the location of recombination breakpoints are indicated over each bar. Persistent homology generators can be used to reconstruct the topology of the tARG, as depicted in (C). Without adding any extra sequences to the sample, the two bars are associated to the same four generators, allowing only to reconstruct the large envelope of the two loops in the tARG. Adding sequences E and F to the sample (represented by blue leaf nodes in (A)) disentangles the generators of the two loops, fully reconstructing the topology of the tARG.}
\label{fig6}
\end{figure}

We illustrate the use and interpretation of barcode ensembles with a simple example, consisting of a sample of 4 genetic sequences with 7 binary characters: 1111001, 1111111, 0000110 and 0000000. Minimal ARGs explaining this sample require two single-crossover recombination events. An ultra-minimal ARG is presented in Fig.~\ref{fig6}A. The most ancestral recombination event involves genetically distant parental gametes, leading to a large loop in the ARG. To the contrary, the most recent recombination event involves genetically close parental gametes, leading to a second small loop in the ARG. These features are captured by the barcode ensemble of the 4 sequences (Fig.~\ref{fig6}B), which consists of two bars, corresponding to the two recombination events. The position of the bars represent the genetic scales associated to the recombination events, with the $2\epsilon_d=5$ ($3$) bar corresponding respectively to the large (small) recombination loop. These death times are good upper bounds for the mutational distance between recombinant sequences in the two genomic intervals associated to each bar (characters 1 to 5, and 6 to 7, respectively). The position of the crossover breakpoints associated to these recombination events is also correctly reproduced. Hence, taking as input the 4 sequences, the barcode ensemble extracts phylogenetic information from ultra-minimal ARGs that explain the sample, without requiring complete reconstruction of the ARGs.

We note here the importance of using the barcode ensemble instead of the ordinary barcode, used in previous phylogenetic applications of persistent homology~\cite{chan2013topology}. In this simple example $b_1=1$, and only one of the two recombination events would have been detected if the ordinary first-homology barcode had been used. The barcode ensemble largely increases the sensitivity to detect recombination events.

We can attempt to reconstruct the tARG of the sample by using persistent homology generators (Fig.~\ref{fig6}C). Whereas there are theorems ensuring the stability of barcodes against small perturbations \cite{cohen2007stability, chazal2009gromov}, the generators of persistent homology identified by TDA strongly depend on the sample, and multiple choices of basis are possible. Hence, the use of generators to reconstruct the tARG is usually limited to small datasets. In this simple example, both bars in the barcode ensemble are generated by the four sampled sequences. Therefore, the reconstructed loop enclosing each recombination event is the same in both cases and corresponds to the large enveloping loop in the ultra-minimal ARG (Fig.~\ref{fig6}A). Adding the internal nodes 1111000 and 1111110 to the sample permits disentangling the generators of the two loops (Fig.~\ref{fig6}C), fully reconstructing the topology of the underlying tARG.

\paragraph{Genetic exchange between two divergent populations.}

\begin{figure}[ht!]
\centering
\includegraphics[scale=0.7]{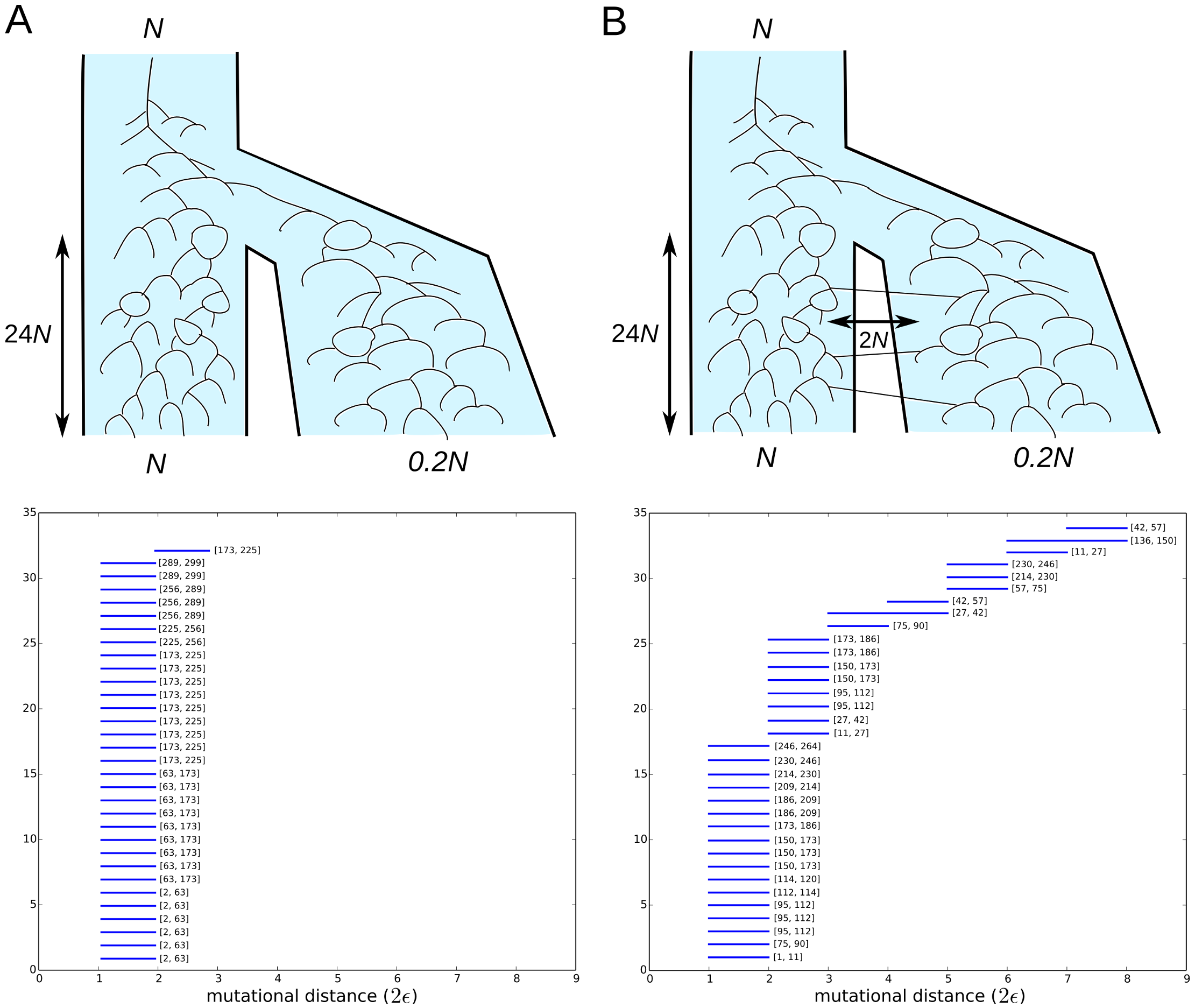}
\caption{{\bf Barcode ensemble of two divergent sexually-reproducing populations}. The case in (A) assumes the two populations are completely isolated. All recombination events present in the barcode ensemble involve genetically close parental gametes. The case in (B) considers a small migration rate between the two populations. Some of the recombination events present in the barcode ensemble involve genetically distant parental strains, leading to larger death times $\epsilon_d$ in the barcode ensemble. The total number of detected recombination events is similar in both cases and uniform across the entire genome. Intervals with the location of the recombination breakpoints are indicated for each recombination event, where positions refer to segregating sites.}
\label{fig8}
\end{figure}

We now consider a more involved example consisting of two sexually-reproducing populations, simulated under the coalescent model with recombination. The two populations diverged $24N$ generations before present. Their effective population sizes are taken to be constant and given by $N$ and $N/5$. We consider two different cases, depicted in Fig.~\ref{fig8}. In the first case (Fig.~\ref{fig8}A), the two populations are completely isolated from each other. In the second case (Fig.~\ref{fig8}B), to the contrary, there is a small migration rate between the two populations. The recombination rate is the same in both cases. Alternatively, in a phylogenetic context, this setting describes the incomplete lineage sorting of two species, with or without the presence of gene flow.

We randomly sampled 250 sequences from the large population and 50 sequences from the small population. The full sample consisted of 300 sequences with 300 segregating sites. We present in Fig.~\ref{fig8} the barcode ensemble for simulated samples without and with migration. The computation took approximately 33 minutes (wall-clock time) in a modern 8-cores desktop computer. Whereas the number of detected recombination events in the tARG, counted by the number bars, is similar in both cases, their genetic scales are very different. Specifically, in the presence of migration the size of some of the loops in the tARG is large, corresponding to migration events followed by a recombination event (Fig.~\ref{fig8}B). This is indicated by the presence of bars with large death time $\epsilon_d$ in the barcode ensemble of the case with migration, corresponding to recombination events with large mutational distances between recombining sequences.

Hence, in this example the barcode ensemble provides rich phylogenetic information that could be hardly obtained by other methods. Methods that attempt to construct a minimal (or nearly-minimal) ARG are computationally inefficient for such large sample sizes \cite{song2003parsimonious, song2005efficient}. Fast bound methods \cite{hudson1985statistical, myers2003bounds, song2005efficient}, on the other hand, do not provide enough phylogenetic information to distinguish between the cases with and without migration, as the total recombination rate is the same in both situations. Sequentially Markov coalescent approaches \cite{rasmussen2014genome} produce an ARG that is far from being minimal but is a good approximation to the maximum likelihood. However, these methods require an underlying coalescent model, with mutation, recombination and population structure parameters given as priors. Finally, algorithms for constructing phylogenetic split networks~\cite{huson2006application} are fast and provide very different outputs in each of the above two cases. However, the interpretation of the output in terms of recombination and migration events is obscure.

\paragraph{Human leukocyte antigen (HLA) locus.} The previous examples serve to illustrate the relation between features of the barcode ensemble of a genetic sample and those of the ultra-minimal ARGs explaining the sample. However, both examples are based on simulated data. We now consider a more realistic example, consisting of 180 phased genotypes from a $\sim$250 kilobase region of the HLA locus of 90 individuals belonging to the Luhya in Webuye, Kenya (LWK) population, sequenced as part of the International HapMap Project \cite{international2010integrating}. In total, the region contains 471 SNPs. Recombination hotspots in this part of the HLA locus have been studied in detail in the past through sperm typing \cite{jeffreys2001intensely} and other high-resolution methods \cite{hinch2011landscape}. This example therefore serves to illustrate the capacity of the barcode ensemble to localize recombination events in realistic situations. With this aim, we also considered 194 phased genotypes from a smaller region (40 kilobase) within the same HLA locus of 97 individuals from the same population, sequenced by the 1,000 Genomes Project Consortium \cite{10002015global}. This additional dataset contained a higher density of SNPs (482 SNPs in total), allowing for a higher resolution in the localization of recombination events. 

\begin{figure}[ht!]
\centering
\includegraphics[scale=0.35]{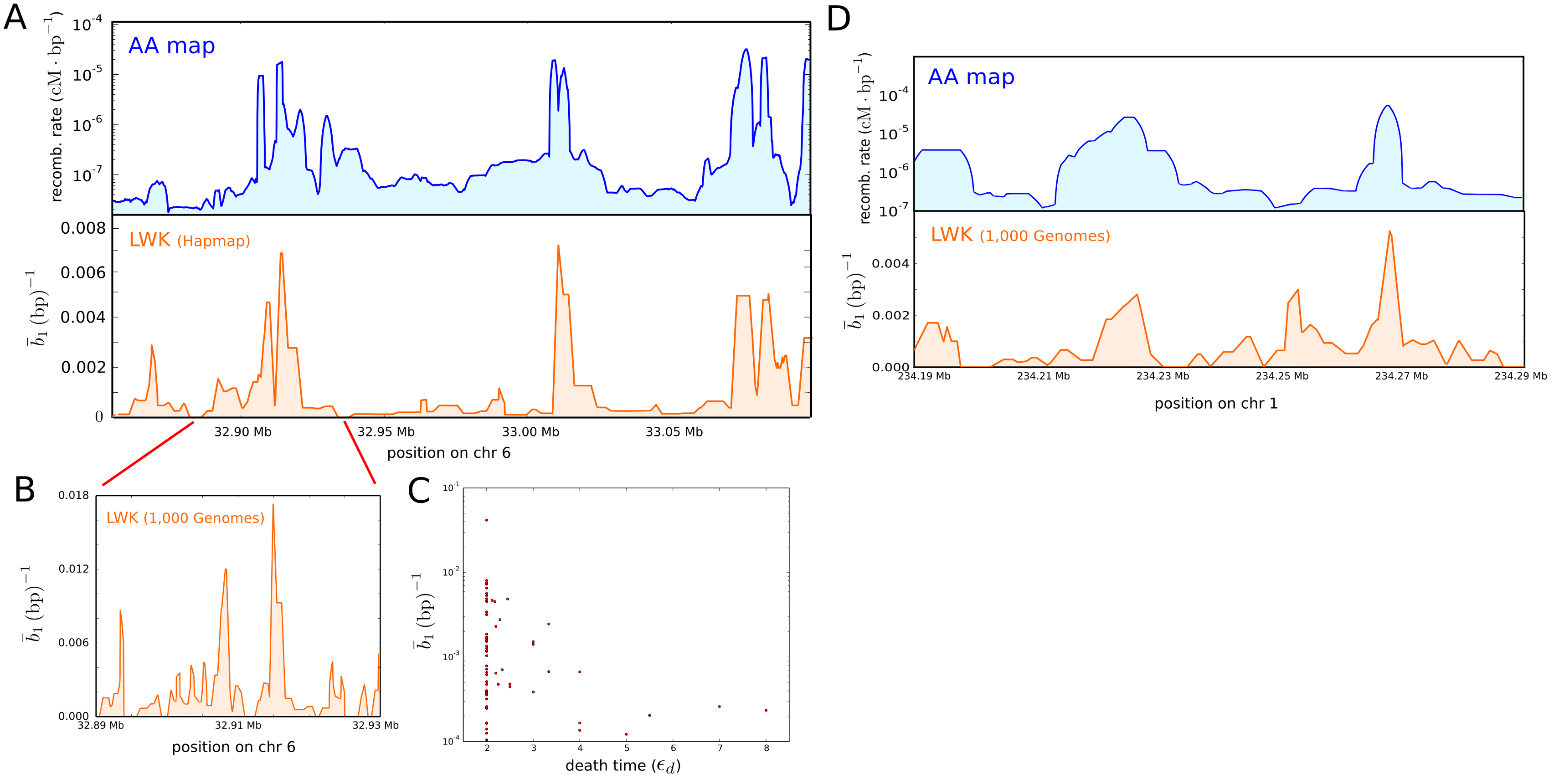}
\caption{{\bf Barcode ensemble across the HLA and MS32 mini-satellite loci of the LWK population}. (A) Recombination rates (top) across a 250 kilobase region of the HLA locus according to the African-American recombination map, based on 30,000 individuals \cite{hinch2011landscape}. The vertical axis is in logarithmic scale. The distribution of recombination events (bottom) detected by the barcode ensemble of a sample of 90 individuals from the LWK population sequenced by the International HapMap Consortium \cite{international2010integrating} is consistent with the observed recombination rates. Note that in neutral models of evolution the number of recombination events in minimal ARGs is roughly expected to grow logarithmically with the recombination rate of the population \cite{hein2004gene}. (B) Distribution of recombination events detected by the barcode ensemble of a sample of 97 individuals from the LWK population, sequenced by the 1,000 Genomes Project Consortium \cite{10002015global}. The higher density of SNPs in this dataset allows for a higher resolution in the localization of recombination events as well as a higher sensitivity. (C) Density of recombination events per nucleotide against their average death time $2\epsilon_d$, for recombination events captured by the barcode ensemble in (A). Each point represents a genomic position for which the barcode ensemble detects recombination. The horizontal axis represents the average death time of the bars in the barcode ensemble that are associated to that genomic position. Events with large $\epsilon_d$, corresponding to recombination events with a large mutational distance between recombining sequences, are mostly associated to regions with low number of recombinations, as expected from neutral models of evolution \cite{hein2004gene}. (D) Recombination rates (top) across a 100 kilobase region near the MS32 mini-satellite locus according to the African-American recombination map \cite{hinch2011landscape}. The vertical axis is in logarithmic scale. The distribution of recombination events (bottom) detected by the barcode ensemble of a sample of 97 individuals from the LWK populations sequenced by the 1,000 Genomes Project Consortium \cite{10002015global} is consistent with the observed recombination rates.}
\label{fig9}
\end{figure}

We used \texttt{TARGet} to compute the first-homology barcode ensembles of the two datasets and analysed the distribution of bars across the HLA locus (Figs.~\ref{fig9}A and \ref{fig9}B). The computation took 19 and 14 minutes (wall-clock time) in a modern 8-cores desktop computer, respectively for the HapMap and 1,000 Genomes datasets. Comparison with the African-American recombination map \cite{hinch2011landscape}, based on more than 2 million crossovers in 30,000 unrelated African-Americans, shows a large degree of consistency between the recombination rates and the genomic position of recombination events detected by the barcode ensembles (Fig.~\ref{fig9}A). The distribution of mutational distances associated to recombination events is qualitatively consistent with coalescent arguments  (Fig.~\ref{fig9}C). In particular, bars with large death time $\epsilon_d$, corresponding to recombination events with a large mutational distance between recombining sequences, are mostly associated to regions of low recombination rate, consistently with the longer coalescence time for these regions \cite{hein2004gene}.

\paragraph{Human MS32 mini-satellite locus.} Similarly, we also considered 194 phased genotypes from a $\sim 100$ kilobase region (416 segregating sites) near the MS32 mini-satellite locus of 97 individuals from the LWK population, sequenced by the 1,000 Genomes Project Consortium \cite{10002015global}. We used \texttt{TARGet} to compute the barcode ensemble and studied the distribution of bars across this genomic region (Fig.~\ref{fig9}D). The computation took 10 minutes (wall-clock time) in a modern 8-cores desktop computer. As in the previous example, the genomic position of the recombination events detected by the barcode ensemble (Fig.~\ref{fig9}D) was consistent with the recombination rate across this region, as determined by the African-American recombination map \cite{hinch2011landscape}.

\paragraph{Darwin's finches.}

Our last example consists of the genetic sequences of 112 Darwin's finches, belonging to 15 different species inhabiting the Gal\'apagos archipielago and Cocos Island \cite{lamichhaney2015evolution}. We aligned and genotyped a 9 megabase scaffold of their genome and, after filtering for high-quality variants, we focussed on a set of 140 SNPs that were homozygous across the 112 samples, thus avoiding potential phasing artefacts. By considering this set, we mostly restrict to very ancestral recombination/gene flow events, close to the origin of radiation from a common ancestor 1.5 million years ago \cite{petren2005comparative}. We used \texttt{TARGet} to obtain the first-homology barcode ensemble of the sample, as well as the partially reconstructed tARG. The computation took 9 minutes (wall-clock time) in a modern 8-cores desktop computer.

\begin{figure}[ht!]
\centering
\includegraphics[scale=0.8]{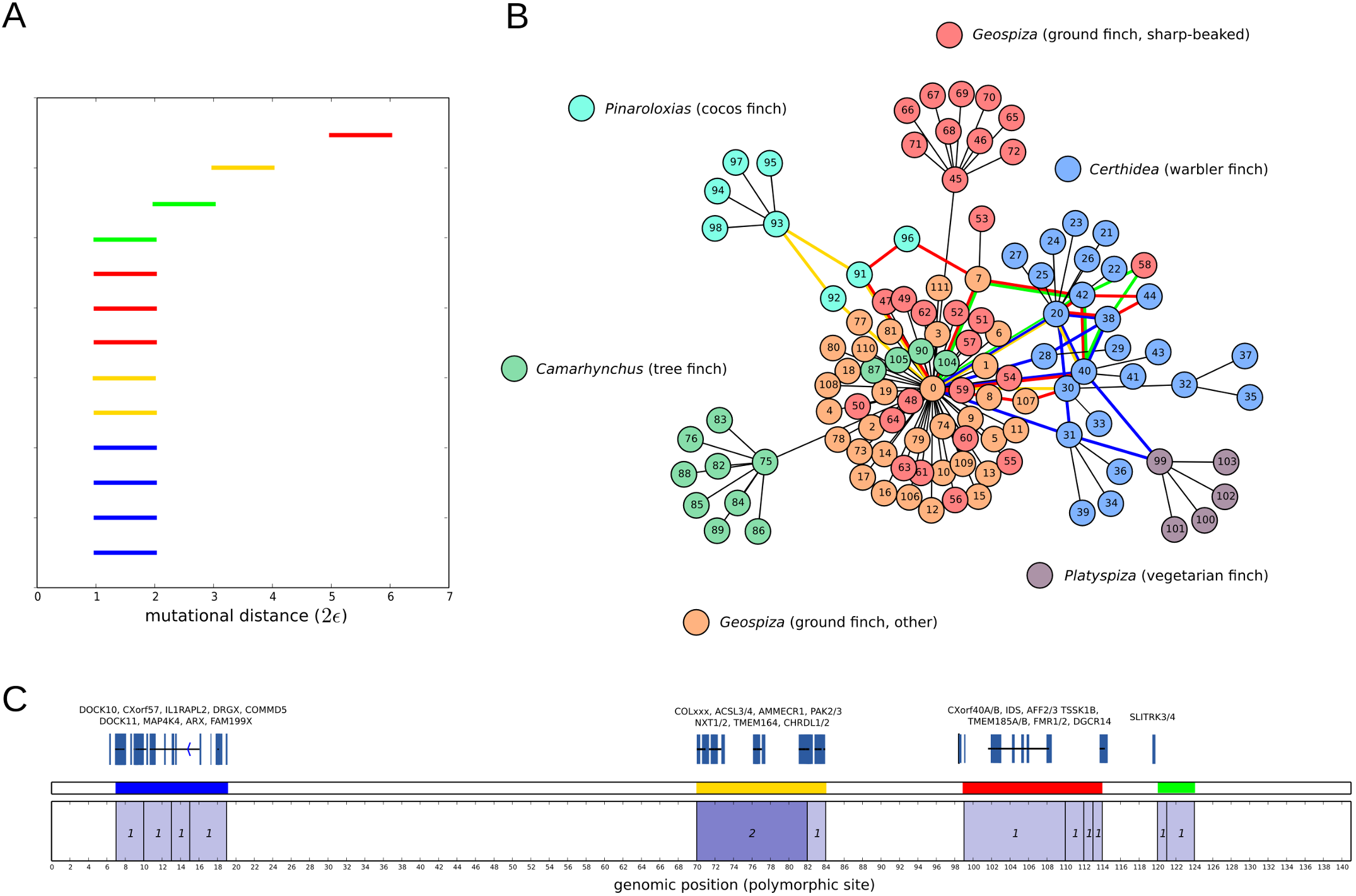}
\caption{{\bf Barcode ensemble and partially reconstructed tARG of a sample of 112 Darwin's finches}. The barcode ensemble is shown in (A), based on 140 homozygous SNPs present in a 9 megabase scaffold. In total, 13 recombination/gene flow events are captured in the barcode ensemble, with different genetic scales. Bars are colored according to the position of the corresponding recombination breakpoint in the genome, as depicted in (C). We also indicate the number of recombination events detected at each genomic interval, as well as some of the orthologous genes present at regions where recombination events are detected. The reconstructed tARG is presented in (B). Loops in the reconstructed tARG are outlined using the same code of colors. We have also included leaf nodes that do not participate in any recombination event, using a nearest neighbour algorithm based on genetic distance. Edge lengths are arbitrary.}
\label{fig7}
\end{figure}

The first-homology barcode ensemble (Fig.~\ref{fig7}A) contains 13 recombination events, mostly involving samples from multiple species and usually including samples from the genus \emph{Certhidea} (Fig.~\ref{fig7}B), the most ancestral lineage among the genera present in the sample \cite{lamichhaney2015evolution}. These results add support to the evidence for genetic introgression found in \cite{lamichhaney2015evolution}. Our analysis also reveals that the crossover breakpoints of these events localize at four different genomic regions within the 9 megabase scaffold that we have considered in this example (Fig.~\ref{fig7}C). 

\subsection{Parameter estimation}

\begin{figure}[ht!]
\centering
\includegraphics[scale=0.8]{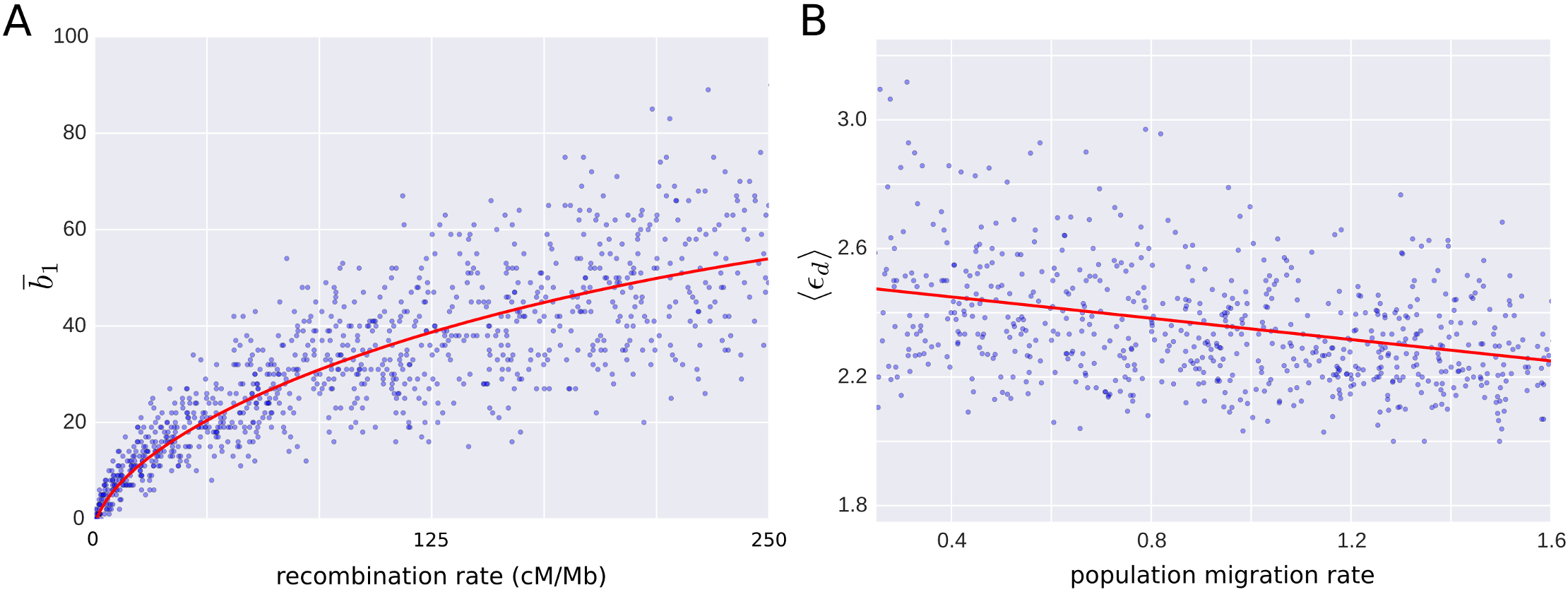}
\caption{{\bf Parameter estimation in models of evolution}. (A) Dependence of $\overline{b}_1$ on the recombination rate parameter for a set of 1,000 simulations of a basic coalescent model. Each simulation consists of 200 sequences, 30 kilobase long. The expected $\overline{b}_1$ of the barcode ensemble grows monotonically with the recombination rate, providing a good measure of the later. The smoothed average is shown in red. (B) Dependence of the average death time, $\langle \epsilon_d\rangle$, on the migration rate of two divergent populations with fixed recombination and variable migration rates, based on 900 simulations. Each simulation consists of 150 sampled sequences, 10 kilobase long. The same structure as in Fig.~\ref{fig8} was considered for the two populations. The expected value of $\langle \epsilon_d\rangle$ decreases monotonically with the migration rate, being informative of the later.}
\label{fig10}
\end{figure}

The examples above illustrate the use and interpretation of barcode ensembles in molecular phylogenetics. As we have discussed, an important feature of topological approaches to phylogenetics is that they inform about most parsimonious evolutionary histories. Being model-independent approaches, they describe minimal sets of events required to explain a sample of sequences, without assuming any probabilistic model of evolution. In some situations, however, we are interested in estimating the parameters of a specific evolutionary model from the observed data (e.g. the recombination rate in a coalescent model with recombination). To that end, barcode ensembles can be taken as summary statistics from which to build parameter estimators. For instance, in Fig. \ref{fig10}A we show the dependence of $\overline{b}_1$ on the recombination rate for a set of 1,000 coalescent model simulations. The expected $\overline{b}_1$ of the barcode ensemble is informative of the recombination rate, growing monotonically with the later. Compared to sequentially Markov coalescent (SMC) approaches for ARG inference \cite{rasmussen2014genome}, $\overline{b}_1$ is strongly correlated with the number of recombinations in SMC ARGs derived from the same set of sequences (Pearson's $r=0.93$, $p<10^{-100}$, Fig. S1). Although the coefficient of variation is $\sim 35\%$ larger for $\overline{b}_1$ (Fig. S1), its computing time is substantially lower ($> 9$ times faster after parallelizing in a modern 8-cores desktop computer, Fig. S1), being a robust approach to coalescent-model recombination rate estimation in large datasets. Furthermore, unlike the number of recombinations in SMC ARGs, $\overline{b}_1$ is unbiassed at small recombination rates, vanishing when the recombination rate is zero (Fig. \ref{fig10}A).

Although recombination rate estimation is a very direct example, the barcode ensemble of a sample of genetic sequences contains other rich phylogenetic information apart from $\overline{b}_1$, which can be used for more complex parameter estimation in structured models of evolution. Consider, for instance, the case of two divergent populations with migration and recombination discussed above. In this model, the average genetic distance between recombining sequences is expected to decrease with the migration rate, as the average time to the most recent common ancestor between foreign and local gametes in a population is shorter.  In figure \ref{fig10}B we show the dependence of the average death time ($\langle \epsilon_d\rangle$) on the migration rate parameter, for the barcode ensembles of a set of 900 coalescent model simulations with fixed recombination and variable migration rates. As expected, $\langle \epsilon_d\rangle$ is informative of the migration rate, decreasing monotonically with the later. It is therefore a good measure for estimating migration rates. Consistently, $\langle \epsilon_d\rangle$ correlates with time to the most recent common ancestor of recombining sequences in SMC ARGs obtained from the same data (Pearson's $r=0.55$, $p<10^{-72}$, Fig. S1). Although the coefficient of variation of $\langle \epsilon_d\rangle$ is $\sim 60\%$ larger (Fig. S1), extracting this type of information from SMC ARGs requires the implementation of a greedy algorithm, substantially increasing the running time ($\sim 8$ times slower in a single core of modern desktop computer, Fig. S1) and therefore limiting its applicability to large datasets. 

These two simple examples illustrate the utility of barcode ensembles for building parameter estimators in specific models of evolution. Importantly, being model-independent, they are robust and flexible tools which can be applied in an infinitely large number of possible evolutionary models.

\section{Discussion}
As the famous title of the essay by Dobzhansky ``Nothing in Biology Makes Sense Except in the Light of Evolution'' underscores,  evolutionary processes are central orchestrating themes in biology. Mutations, recombinations and other evolutionary processes get imprinted into genomes through selection, reflecting the accumulated history giving rise to an organism. Phylogenetics try to reconstruct the evolutionary history through the comparison of genomes of related organisms. In addition to reporting relationships and elucidating particular histories, one would like to understand and quantify how different evolutionary processes have occurred. The identification and quantification of evolutionary processes can be challenging due to the lack of a well-established universal framework to capture evolutionary relationships beyond trees. In addition, robust statistical inference needs to exploit the large number of genomes that are now becoming available, aggravating the computational burden and obscuring interpretations. Ideally, we would like to have a biologically interpretable framework able to quantify different evolutionary processes by analyzing large numbers of genomes.     

In this paper we have proposed a few steps in this direction. We have extended the notion of barcodes in persistent homology to identify the genetic scale and number of recombination events. We have shown that, by correctly studying persistent homology in subsets of segregating sites, it is possible to characterize the genomic regions where recombination takes place and identify the gametes involved in particular recombination events. The persistent homology barcodes derived from each of these sets can be structured as a ``barcode ensemble'' where each bar captures a recombination event. Barcode ensembles can be interpreted as counting and quantifying the scale of recombination events in a variation of Ancestral Recombination Graphs (ARGs). Topological ARGs represent a summary of potential recombination histories that can explain the data. The method proposed, \texttt{TARGet}, is scalable to hundreds of genomes. As an alternative to some phylogenetic networks, barcode ensembles provide robust quantification of events, the distribution of genetic scales, computational scalability and interpretative graphs.

Barcode ensembles are versatile in that they do not assume any specific model of evolution, providing explicit, interpretable summaries of the minimal set of recombination events required to explain a sample of genetic sequences. Here we have illustrated their use in several practical cases. However, the range of possible applications is unlimited. In some cases, it may be convenient to perform minor modifications to the approach described here. For instance, although in our exposition we have only made use of Hamming distance and binary sequences, the main concepts we have presented extend straightforwardly to other genetic distances. The use of these metrics can be particularly useful in cases with rapidly diverging samples or substantial mutational biases. In other cases, information about the ancestral and derived alleles for each character in the sample may be available. Although tARGs have no natural directionality, the inclusion of the ancestral sequence in the original sample may lead in those cases to more stringent bounds on $\overline{R}_{\rm min}$, similarly to what occurs with other approaches to recombination inference \cite{gusfield2014recombinatorics}. Finally, more efficient integer linear programming algorithms, like the one of~\cite{song2005efficient}, could in principle be also generalized to the computation of barcode ensembles.

\section{Methods}
\subsection*{First-homology barcode ensemble}
{\small We extended the construction of ref.~\cite{myers2003bounds} to persistent homology barcodes. From a geometric perspective, this corresponds to projecting the original space on sets of mutually orthogonal hyperplanes in the ambient hypercube, and computing persistent homology in each of those projections. For that aim, we need to establish an ordering relation on barcodes. Being sets of intervals, it is natural to take the maximum of two barcodes to be given by the one with largest $L^0$-norm, namely largest $b_1$. If both barcodes have the same $L^0$-norm, we may successively compare other norms (e.g. other $L^p$-norms), until the tie is broken or, otherwise, one of the two barcodes is arbitrarily chosen. The algorithm of \cite{myers2003bounds} is then generalized to persistent homology barcodes as follows:
\begin{enumerate}
\item Let $\mathcal{B}_{ik}$ be the first-homology barcode of the sequences that result from the $i$-th to $k$-th characters in $\mathcal{S}$. Set $\mathcal{R}_{ij} = 0$ and $k = 2$.
\item For $j=1,\ldots,k-1$, set $\mathcal{R}_{jk} = \textrm{max}\{\mathcal{R}_{ji} \cup \mathcal{B}_{ik}\ :\ i = j, . . . , k - 1\}$
\item If $k < m$, increment $k$ by 1 and go to step 2.
\end{enumerate}
The barcode ensemble of $\mathcal{S}$ is the union barcode $\mathcal{R}_{1m}$ that results from this algorithm. 

We implemented the algorithm in a publicly available multi-threaded software, \texttt{TARGet}, which is distributed under the GNU General Public License (GPL v3). The application is fully written in Python 2.7, and relies on Dionysus C++ library for persistent homology computations (\texttt{http://www.mrzv.org/software/dionysus}). Since considering all possible sequence partitions is unnecessary and computationally infeasible in most cases, we follow the strategy of ref.~\cite{myers2003bounds} and allow the user to limit the number of partitions by the maximum number of segregating characters within each subset of $\mathcal{S}$ (specified by the command line option \texttt{-s}), and by the maximum distance between segregating characters in the subset (specified by the command line option \texttt{-w}). In addition, we also allow the user to exclude from $\mathcal{S}$ segregating characters that are compatible (namely, that satisfy the Hudson-Kaplan four-gamete test \cite{hudson1985statistical}) with all the other characters in $\mathcal{S}$ (specified by the command line option \texttt{-e}). For each genomic interval, a filtration of Vietoris-Rips complexes is constructed using Hamming distance and the persistent first-homology group is computed over $\mathbf{Z}_2$.}

\subsection*{\small Population genetics simulations}
{\small We performed 4,000 simulations of a sample of 40 sequences with 12 segregating sites, using the software \texttt{ARGweaver} \cite{rasmussen2014genome}. The population was simulated using a coalescent infinite sites model with recombination. The population-scaled recombination rate, $\rho$, was randomly generated in each simulation, taking values from a uniform distribution between 0 and 110. For each simulated sample, Myers and Griffiths lower bound $R_{\rm MG}\leq R_{\rm min}$ was computed using the software \texttt{RecMin} \cite{myers2003bounds}, with parameters \texttt{-s 12 -w 12}. Lower bounds $\bar b_1\leq \overline{R}_{\rm min}$ were computed using our application \texttt{TARGet}, with parameters \texttt{-s 12 -w 12}.

To study the dependence of $\overline{b}_1$ on the recombination rate parameter in coalescent models, we performed 1,000 simulations of a sample of 200 sequences. The population-scaled recombination rate, $\rho$, was randomly generated in each simulation, taking values between 0 and 216. For each simulated sample, \texttt{TARGet} was run with parameters \texttt{-s 11 -w 11}, and \texttt{ARGweaver}'s tool \texttt{arg-sample} was run with parameters \texttt{-m 7e-9 -n 400 --sample-step 10}, discarding the first 200 iterations. 

Samples of genetic exchange between two divergent populations were simulated using the software \texttt{ms} \cite{hudson2002generating}, using the commands

\texttt{ms 300 1 -s 300 -r 40 10000 -I 2 250 50 -ej 6.0 1 2 -n 2 0.2 -m 1 2 0.5} 

\noindent and,

\texttt{ms 300 1 -s 300 -r 40 10000 -I 2 250 50 -ej 6.0 1 2 -n 2 0.2}

\noindent respectively for the cases with and without migration. The barcode ensemble of each sample was computed using \texttt{TARGet} with parameters \texttt{-s 12 -w 14 -e}.

To study the dependence of $\langle\epsilon_d\rangle$ on the migration rate in this scenario, we performed 900 simulations using the software \texttt{ms} \cite{hudson2002generating} and \texttt{seq-gen} \cite{rambaut1997seq}, with the commands

\texttt{ms 150 1 -T -r 60 10000 -I 2 125 25 -ej 6.0 1 2 -n 2 0.2 -m 1 2 X}

\noindent and

\texttt{seq-gen -mHKY -l 10000 -s 0.004 -p 50000}

\noindent where the migration rate \texttt{X} in the first command takes random values from a uniform distribution between 0 and 2. For each simulated sample, \texttt{TARGet} was run with parameters \texttt{-w 8 -s 8}, and \texttt{arg-sample} was run with parameters \texttt{-m 1e-7 -n 400 -r 1.5e-7}. We extracted from SMC ARGs the time to the most recent common ancestor of recombining sequences using a greedy algorithm that searches for the shortest non-zero path connecting the two sequences.}

\subsection*{\small HLA and MS32 loci}

{\small We downloaded phased genotype data from HapMap phase III \cite{international2010integrating}, corresponding to all SNPs of LWK population between rs6457661 and rs3129301 in chromosome 6. We also downloaded phased genotype data from 1,000 Genomes Project \cite{10002012integrated}, corresponding to all SNPs of LWK population between positions 32,887,978 and 32,927,978 of chromosome 6, and half of the SNPs of LWK population between positions 234,190,031 and 234,291,193 of chromosome 1. All coordinates refer to human assembly hg18. The barcode ensemble of each dataset was computed using \texttt{TARGet} with parameters \texttt{-s 12 -w 12}.}

\subsection*{\small Darwin's finches genotyping}
{\small Raw paired-end reads from 112 Darwin finches \cite{lamichhaney2015evolution} were obtained from SRA archive (accession number PRJNA263122) and aligned against the consensus sequence of \emph{Geospiza Fortis}, version GeoFor\_1.0/geoFor1, scaffold JH739904. We followed essentially the same procedure than that of ref.~ \cite{lamichhaney2015evolution} for the alignment, SNP calling, genotyping and filtering. In short, the alignment was performed with Burrows-Wheeler aligner (BWA) \cite{li2009fast}, version \texttt{0.7.5}, using BWA-MEM algorithm and default parameters. PCR duplicates were marked using Picard tools (\texttt{http://picard.sourceforge.net/}). Indel realignment, SNP discovery and simultaneous genotyping across the 112 samples was performed using Genome Analysis Toolkit (GATK) \cite{mckenna2010genome}, following GATK best practice recommendations \cite{depristo2011framework}. SNP calls were filtered by keeping variants with SNP quality $>$ 100, total depth of coverage $>$ 117 and $<$ 1750, ratio between SNP quality and depth of coverage $>$ 2, Fisher strand bias $<$ 60, mapping quality $>$ 50, mapping quality rank $>$ -4 and read position rank sum $>$ -2. In total, 13,980 variant positions passed these filters. To avoid phasing errors, we only considered SNPs that were homozygous across the 120 samples. The resulting genotypes were processed with \texttt{TARGet} for barcode ensemble computation, using the options \texttt{-s 14 -w 14}.}

\section*{Acknowledgments}
We thank C.~Solis-Lemus, C.~Ane, D.~S.~Rosenbloom, K.~Emmett, M.~Lesnick, S.~Zairis and V.~Aleksandrov for useful discussions and comments.

\newpage
\begin{figure}[ht]
\centering
\includegraphics[scale=0.8]{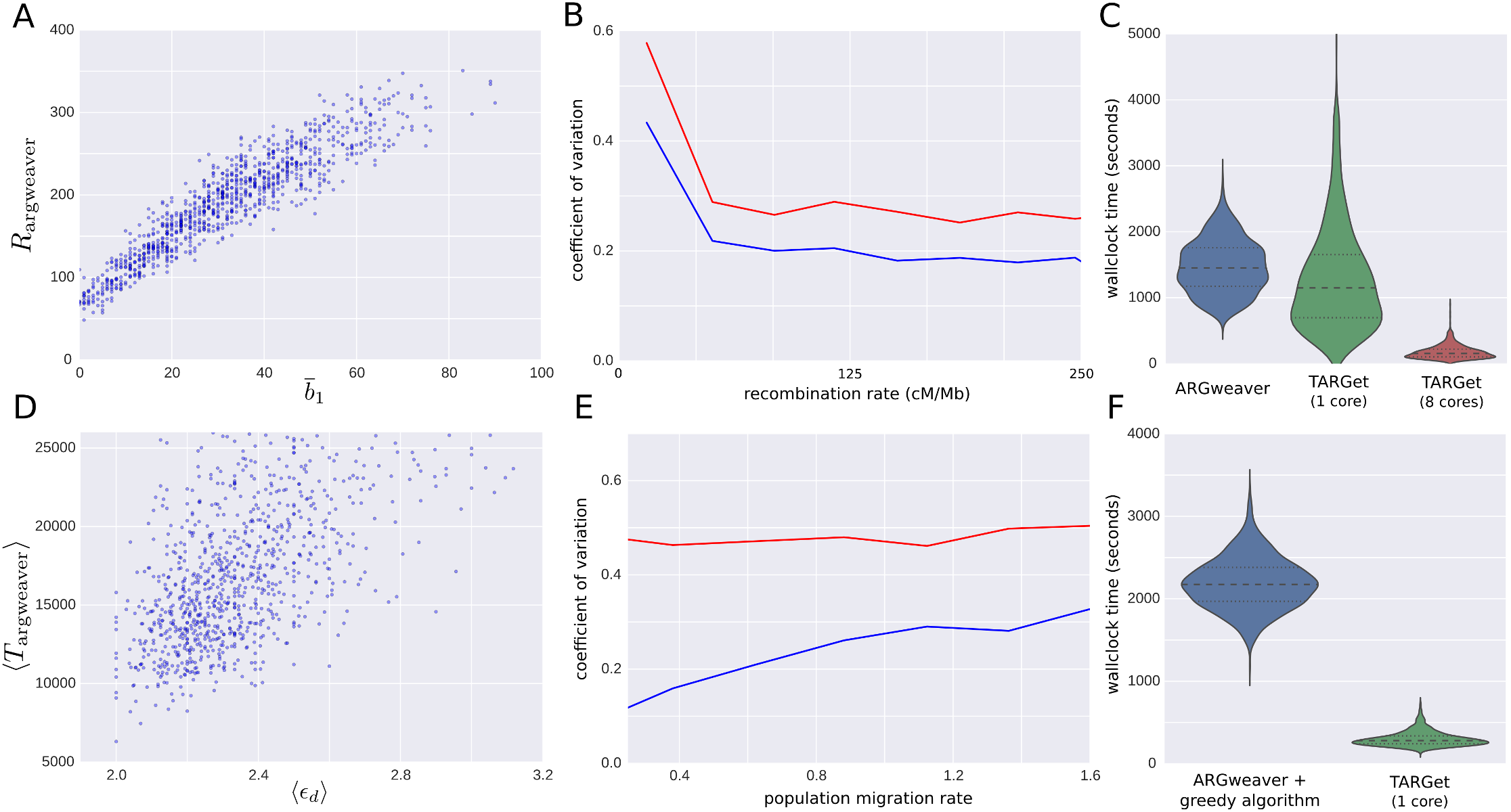}
\end{figure}
\noindent Figure S1. {\bf Comparison to SMC approaches to ARG inference.} (A) The number of recombination events in SMC ARGs \cite{rasmussen2014genome}, $R_{\rm argweaver}$, plotted against the number of bars in the barcode ensemble, $\overline{b}_1$. Both quantities are strongly correlated (Pearson's $r=0.93$, $p<10^{-100}$). Plot based on 1,000 coalescent model simulations of a sample of 200 sequences. (B) Coefficient of variation of $R_{\rm argweaver}$ (blue) and $\overline{b}_1$ (red) as a function of the recombination rate. (C) Distribution of wall-clock running times for the simulations in (A). (D) Average time to the most recent common ancestor of recombining sequences in SMC ARGs, $\langle T_{\rm argweaver}\rangle$, plotted against the average death time of bars in the barcode ensemble, $\langle \epsilon_d\rangle$, for two divergent populations with recombination and migration. Both quantities are largely correlated (Pearson's $r=0.55$, $p<10^{-72}$). Plot based on 900 simulations of a sample of 150 sequences. (E) Coefficient of variation of $\langle T_{\rm argweaver}\rangle$ (blue) and $\langle \epsilon_d\rangle$ (red) as a function of the migration rate. (F) Distribution of wall-clock running times for the simulations in (D). \texttt{ARGweaver} running times also include the time required to extract $\langle T_{\rm argweaver}\rangle$ from SMC ARGs.

\end{document}